\documentclass[journal]{IEEEtran}
\usepackage{times,amsmath,color,amssymb,graphicx,epsfig,cite,psfrag,subfigure,algorithm,balance}
\usepackage{amsfonts,pifont,enumerate,cases}
\usepackage{mathrsfs}
\usepackage[table]{xcolor}
\usepackage{verbatim}
\usepackage{bm}
\usepackage{url}

\DeclareMathOperator*{\argmin}{argmin}

\newcommand{\vect}{\textup{vec}}

\begin{document}
\title{An Overview of Enhanced Massive MIMO with Array Signal Processing Techniques}
\author{Mingjin Wang, Feifei Gao, Shi Jin,  and Hai Lin

\thanks{M. Wang and F. Gao (\emph{Corresponding author}) are with the Department of Automation, Tsinghua University, Beijing
100084, China, also with the Institute for Artificial Intelligence, Tsinghua
University (THUAI), Beijing 100084, P. R. China, and also with the Beijing
National Research Center for Information Science and Technology (BNRist),
Beijing 100084, P. R. China (e-mail: wmj17@mails.tsinghua.edu.cn, feifeigao@ieee.org).}
\thanks{S. Jin is with the National Communications Research Laboratory, Southeast University, Nanjing 210096, P. R. China (email: jinshi@seu.edu.cn).}
\thanks{H. Lin is with the Department of Electrical and Information Systems,
Graduate School of Engineering, Osaka Prefecture University, Sakai, Osaka, 5998531, Japan (email: hai.lin@ieee.org).}
}
\maketitle
\thispagestyle{empty}
\vspace{-18mm}
\begin{abstract}
In the past ten years,  there have been tremendous research progresses on massive MIMO systems, most of which stand from the communications viewpoint. A new trend of investigating massive MIMO, especially for the sparse scenario like millimeter wave (mmWave) transmission, is to re-build the transceiver design from  array signal processing viewpoint that could  deeply exploit the half-wavelength array and
provide enhanced performances in many aspects. For example, the high dimensional channel could be decomposed into small amount of physical parameters, e.g., angle of arrival (AoA), angle of departure (AoD), multi-path delay, Doppler shift, etc. As
a consequence, transceiver techniques like synchronization, channel estimation, beamforming, precoding, multi-user access, etc., can be re-shaped with these physical parameters, as opposed to those designed directly with  channel state information (CSI). Interestingly,  parameters like AoA/AoD and multi-path  delay are frequency insensitive and thus  can be used to guide the downlink  transmission from uplink training even for FDD systems.  Moreover, some phenomena of
massive MIMO that were vaguely revealed previously can be better explained now with array signal processing, e.g., the beam squint effect. In all, the target of this paper is to present an overview of recent progress on merging array signal processing  into  massive MIMO communications as well as its promising future directions.
\end{abstract}


\section{Introduction}
The fifth generation (5G) wireless communications demands for a substantial increase in transmission throughput and network coverage in order to support a broad range of emerging applications, such as smart phones, multimedia, social networks, internet gaming, etc. One promising physical layer technology of 5G is the massive multiple-input multiple-output (MIMO) that scales up conventional MIMO by several orders of magnitude  and advocates the use of a few hundred or thousand antennas at the base station (BS) to greatly increase the system capacity over available time-frequency resources \cite{MIMO1,MIMO2,MIMO3}. It promises to reap all the benefits of conventional MIMO, and hence significantly improves the spectrum and power efficiencies of wireless communications.
In recent years, there has been tremendous theoretical research on massive MIMO systems, including capacity analysis \cite{MIMO-capcity}, channel estimation \cite{MIMO-CE}, synchronization \cite{MIMO-sync}, beamforming technique \cite{MIMO-bf}, user scheduling \cite{MIMO-user}, spectral efficiency \cite{MIMO-efficiency}, etc.

Practically, massive MIMO  system  chooses the antenna spacing as half-wavelength to maintain an implementable array aperture, which is quite  different from the convention where the antennas are often placed further away from each other to achieve the spatial diversity. With half-wavelength arrays, nevertheless, the array signal processing techniques for phased array in Radar and Sonar applications \cite{capon,music,esprit,signal-pro}  can possibly be  applied to enhance the quality of wireless data communications.

Array signal processing  has long been  used in military applications to extract the angles of Radar targets or to formulate narrow beams for jamming/anti-jamming  \cite{radar}. The difference from wireless communications is that the information contained in signals is not cared but rather the AoA that represents the target's position. In fact, the famous term \emph{beamforming}  was originated from array signal processing that means physically formulating an  electromagnetic beam towards the target. Later on, the term \emph{beamforming} was replanted in wireless communications to represent the weights of multiple antennas that can maximize the signal-to-noise ratio (SNR) of a user while does not necessarily formulate a physical beam over the space.

The first trial to combine the  area of  array signal processing and  wireless communications  was probably the smart antennas \cite {smart-antenna} that arose around the mid-to-late 1990's when the ArrayComm deployed  angle-based spatial division multiple access (SDMA) base station (BS) in both Japan and Australia.
Unfortunately, this first trial was not that successful due to two-fold reasons: (i) Unlike in radar environment, microwave wireless communications have too many multi-paths from reflection and scattering such that the AoAs of incoming paths are not distinguishable; (ii) The number of antennas in conventional MIMO is small, say 4 or 8, which cannot support accurate  AoA estimation while would bring large side lobes when formulating the physical beams.

This time, owing to the large antenna size, massive MIMO is well positioned for successfully deploying array signal processing techniques since the spatial information of users can be precisely identified and very narrow beams can be formulated to multiplex different users. Meanwhile, the tendency to high frequency communications, e.g., millimeter wave (mmWave) would face gigantic  path loss such that the total number of  effective paths is small. Particularly, technical terms in array signal processing (e.g., angle of arrival/departure - AoA/AoD), can be conceptualized in the terms of wireless communications, such as channel estimation, hybrid beamforming, interference control, multiple access scheme, etc., which suggests an exciting research direction of leveraging advanced array signal processing techniques to aid wireless communications. In fact, since massive MIMO is inherently built upon the multi-antenna array, it is expected that the mature array signal processing technologies could be exploited deeply into this architecture and provide more reliable designs as well as the enhanced performance compared to those directly obtained from the communications viewpoint.

In all, the purpose of this article is to provide an overview of
the state-of-the-art results that apply array signal processing techniques for massive MIMO
communication systems.
We intend not to duplicate those  transceiver techniques that have already been introduced in \cite{5G-survey,Liye-survey,mmwave-survey,Heath-survey} from communications viewpoint  but rather focus on new designs from array signal processing viewpoint.

The paper is organized as follows. Section II presents the channel models of  massive MIMO communication systems  from array signal processing viewpoint where the channels can be divided into spatial/frequency and narrowband/wideband categories. A detailed overview of parameter estimation, both blind and training based, is provided in Section~III, where it is pointed out that parameters like AoA/AoD, multi-path delay, and Doppler shift are widely  frequency insensitive and is helpful for downlink channel estimation even in FDD systems.
Section~IV discusses the synchronization issue in massive MIMO systems.
Section~V describes angle-domain hybrid precoding and interference control architectures.
We introduce  orthogonal time and frequency space (OTFS) modulation in massive MIMO systems with  \emph{angle-delay-Doppler} 3D structured channels in Section~VI.
With physical parameters, we present in Section~VII  three new multiple access technologies and illustrate their relationship with conventional time/frequency/spatial division multiple access schemes.
Section~VIII  investigates some artificial intelligence (AI)-based signal processing techniques for parametric
channels in massive MIMO systems.
Section~IX concludes the paper and outline future directions to further merge array signal processing  with massive MIMO communications.

\textbf{Notations:}
Throughout this paper, vectors and matrices are denoted by boldface lower-case and upper-case letters, respectively;
the transpose,  Hermitian, inverse, and pseudo-inverse of the matrix ${\mathbf A}$ are denoted
by ${\mathbf A}^T$, ${\mathbf A}^H$,${\bm A}^{-1}$ and ${\mathbf A}^{\dagger}$, respectively;
$\vect(\bm A)$ represents column-major vectorization of the matrix $\bm A$, i.e., the operation of stacking the columns of matrix $\bm A$ to form a vector;
$\|\bm h\|_0$, $\|\bm h\|_1$, and $\|\bm h\|_2$ are the  $\ell_0$-norm,  $\ell_1$-norm and $\ell_2$-norm of vector $\bm h$, respectively;
$*$ denotes the convolution operator;
$\mathbb R$ represents the set of real numbers;
$\mathbb C$ represents the set of complex numbers;
$\mathcal R(\cdot)$ is the real part of a complex number, vector or matrix.

\section{Channel  Characterise of Large Antenna Array}
Let us first briefly present how the wireless communications channels can be  modeled from  array signal processing viewpoint.
For ease of illustration, we consider the massive MIMO systems with $M\gg 1$ receive antennas at the BS in the form of a uniform linear array (ULA), while the related discussion can be readily extended to planar array cases.
The carrier frequency is denoted by $f_c$, and the carrier wavelength is denoted by $\lambda_c\triangleq c/f_c$, where $c$ is the speed of light.  The antenna spacing is denoted by $d\triangleq\lambda_c/2$. Meanwhile, we assume  the communications system has symbol period $T$ and bandwidth $W$.

\subsection{Channel  Modeling}
Suppose a continuous-time passband signal $\mathcal{R}\{s(t)e^{j2\pi f_ct}\}$ is sent from far field user and arrives at BS via $P$ multi-paths  (uplink). Each path can be characterized by four parameters: the AoA $\theta_p \in[-\pi/2,\pi/2)$, the passband path gain $\tilde{\beta}_p \in \mathbb R$, the propagation delay to the first receive antenna $\tau_p$, and the Doppler shift $\nu_{p}\in[-\nu_{\max}/2,\nu_{\max}/2)$, respectively. Here, $\tau_{\max}$ and $\nu_{\max}$ are termed as the delay spread and (two-sided) Doppler spread of the channel, respectively.
Moreover, the $p$th path arrives at the $m$th receive antenna  $\Delta {\tau}_{m,p}= \frac{(m-1)d\sin \theta_p}{c}$  seconds after it arrives at the first receive antenna, where $c$ is the speed of light.
The received passband continuous-time  signal at the $m$th antenna can be described by   {\cite{h-model,h-time-vary}}\footnote{When $P$ goes to infinite, the sum function can be replaced by integration, i.e., the incident angles are in the continuous domain.}
\begin{align}
  \tilde{y}_m(t)\!=\!\!\!\sum_{p=1}^P\!\mathcal{R}\!\left\{\tilde{\beta}_{p} s(t\!-\!\!(\tau_p\!\!+\!\Delta {\tau}_{m,p}\!))e^{j2\pi (f_c\!+\nu_{p})(t-(\tau_p\!+\!\Delta {\tau}_{m,p}))}\!\!\right\}.
\end{align}
Removing the carrier frequency  $e^{j2\pi f_ct}$, the baseband received signal is
\begin{align}\label{yt}
 y_m(t)&=\sum_{p=1}^{P}\beta_{p}  e^{-j2\pi (f_c+\nu_{p})\Delta{\tau}_{m,p}}e^{j2\pi \nu_{p}t}s\big(t-\tau_{p}-\Delta{\tau}_{m,p}\big)\notag\\
 &=\int h_m(t,\tau) s(t-\tau)d\tau,
\end{align}
where $\beta_p=\tilde{\beta}_{p}e^{-j2\pi (f_c+\nu_{p})\tau_{p}}$ is the equivalent baseband path gain, while the last equation represents the linear time varying filter with the equivalent time varying channel impulse response (CIR)
\begin{align}
h_m(t,\tau)\!\!=\!\!\sum_{p=1}^{P}\beta_{p}  e^{-j2\pi (f_c+\nu_{p})\Delta{\tau}_{m,p}}e^{j2\pi \nu_{p}t}\delta(\tau\!\!-\!\tau_p-\!\!\Delta{\tau}_{m,p}).\label{eq:gao1}
\end{align}
Taking the Fourier Transform of $h_m(t,\tau)$ over $\tau$, we obtain the frequency domain time varying channel response at the $m$th antenna as
\begin{align}
h_m(t,f)\!\!=\!\!\sum_{p=1}^{P}\!\beta_{p}  e^{-j2\pi (f_c+\nu_{p})\Delta{\tau}_{m,p}}e^{j2\pi \nu_{p}t}e^{-j2\pi f(\tau_p+\Delta{\tau}_{m,p})}.
\end{align}
Stacking the channel from all $M$ antennas,  the  frequency domain time varying  channel response vector of the whole antenna array can be expressed as \cite{h-model,h-time-vary}:
\begin{align}\label{htf-vector}
\mathbf{h}(t,f)=\sum_{p=1}^{P}\beta_{p}   \bm a (\theta_p,f,\nu_{p})  e^{-j2\pi f \tau_p} e^{j2\pi \nu_{p}t} ,
\end{align}
where $\bm a(\theta_p,f,\nu_{p})$ is the $M\times 1$  wideband array steering vector:
\begin{align}\label{steering-narrow}
\bm a(\theta_p,f,\nu_{p})=\big[1,&e^{-j2\pi (1+\frac{f+\nu_{p}}{f_c})\frac{d\sin\theta_{p}}{\lambda_c}},\cdots,\notag \\
 &\kern 30pt e^{-j2\pi (1+\frac{f+\nu_{p}}{f_c})\frac{(M-1)d\sin\theta_{p}}{\lambda_c}}\big]^T.
\end{align}

\begin{figure}[t]
  \centering
  \vspace{-0.5cm}
  \includegraphics[scale=0.35]{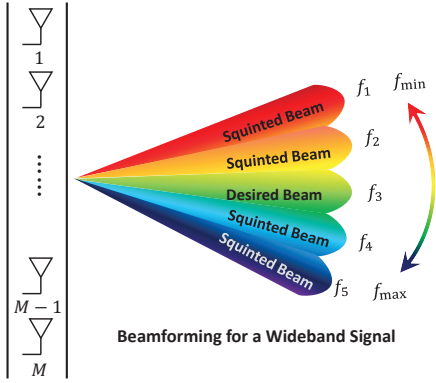}
  \caption{Illustration of beam squint effect in large-scale antenna array}
  \label{beamsquint}
\end{figure}

In  orthogonal frequency division multiplexing (OFDM) system,  the signal bandwidth is divided into $N$ subcarriers. According to \eqref{steering-narrow} the steering vectors at different subcarriers are affected by  their frequencies, i.e., the steering vectors are different for different subcarriers even with the same AoAs. This phenomenon is known as  the \emph{beam squint effect} \cite{bs1,bs2,bs-dai,bs3}, as shown in Fig.~\ref{beamsquint}. \footnote{The existing papers on beam squint effect  have been hypothesized directly from frequency domain of the broadband communications \cite{Heath-survey,bs1,bs2,bs-dai}, which did not reveal the fundamental reasons of  this  problem.}

\subsection{Spatial Narrowband Modeling }
When the signal bandwidth $W$  or the number of antennas $M$ is small, such that $\Delta{\tau}_{m,p}\ll T$,  for $\forall p,  p \in \{1,\cdots,P\},\forall m, m \in \{1,\cdots,M\}$ (the propagation delay across the array is negligible compared to symbol period),  the signals on all antennas can be approximated by  $s(t-\tau_{p}-\Delta{\tau}_{p,m})\approx s(t-\tau_{p})$; namely,  different receive antennas effectively observe the synchronized waveform. This case is called as \emph{narrowband} modeling  for array signal processing but should actually be more accurately named as \emph{spatial narrowband} modeling in wireless communications.
In this case, one can stack the time domain baseband received signal  from all $M$ antennas and obtain the conventional  massive MIMO model as  \cite{MIMO-capcity,MIMO-CE,MIMO-sync,MIMO-bf,MIMO-F1,MIMO-F2,MIMO-F3,MIMO-F5,MIMO-F6}:
\begin{align}
\mathbf{y}(t)\!=\![y_1(t),y_2(t),\ldots, y_M(t)]^T \!\!=\mathbf{h}(t,\tau)*s\big(t-\tau_{p}\big),
\end{align}
and the equivalent time domain MIMO channel vector is\footnote{Time domain MIMO channel vector would lose the steering vector type structure for the general modeling (\ref{yt}) due to the unsynchronized signal waveform at different antennas.}
\begin{align}\label{httao}
\mathbf{h}(t,\tau)=\sum_{p=1}^{P}\beta_{p}\mathbf{a}(\theta_p, 0,\nu_{p})e^{j2\pi \nu_{p}t}\delta(\tau-\tau_p).
\end{align}
It can be easily checked that $\bm{h}(t,f)$ does not have beam squint effect anymore.
\subsection{Frequency Narrowband  Modeling}
When the bandwidth $W$ is small or the multi-path delay is small such that $\tau_{p}\ll T$,  for $\forall p,  p \in \{1,\cdots,P\}$ (the multi-path delay  is negligible compared to symbol period), 
\eqref{htf-vector} reduces to
\begin{align}\label{htf-narrow-frequency}
\bm {h}(t,f)=\sum_{p=1}^{P}\beta_{p}   \bm a (\theta_p,f,\nu_{p}) e^{j2\pi \nu_{p}t},
\end{align}
and the case is called \emph{frequency narrowband} modeling. Interestingly,  though the frequency dependent factor $ e^{-j2\pi f \tau_p}$ disappears (this is how the conventional frequency flat fading establishes for small MIMO), the channel response vector $\bm {h}(t,f)$ is still varying with frequency $f$ due to the beam squint effect. In this case, the inter symbol interference (ISI) in time domain still exists and the OFDM modulation with certain amount of cyclic prefix (CP) should, again, be applied \cite{bolei-bs}.

\subsection{Spatial and Frequency Narrowband  Modeling}
If both $\tau_{p}W\ll 1$ and $\Delta{\tau}_{m,p}W\ll 1$ hold,  for $\forall p,  p \in \{1,\cdots,P\},\forall m, m \in \{1,\cdots,M\}$, then the channel response vector reduces to
\begin{align}\label{htf-narrow-vector}
 \bm h(t)=\sum_{p=1}^{P}  \beta_{p} \bm a(\phi_{p},0,\nu_p) e^{j2\pi \nu_{p}t},
\end{align}
which yields the same frequency domain channels for all subcarriers and truly presents to the  flat fading environment. This case is then named as
\emph{spatial- and frequency- narrowband} modeling. In comparison, the very general channel
model (\ref{eq:gao1}) and (\ref{htf-vector}) for massive MIMO are named as \emph{spatial- and frequency-wideband} model or \emph{dual-wideband} model \cite{bolei-bs, bolei-bs1}.

\subsection{ Summary of Wideband-Narrowband Channel Modeling}
According to the physical parameters, e.g.,  AoA, delay, and Doppler shifts, the channel of massive MIMO system can be classified into different
categories, as shown in  Tab. \ref{table}:
\begin{table}[H]
  \centering
	\begin{tabular}{|r||l|l|l|}
		\hline
		Channel Classification & $W \Delta{\tau}_{M,p}$  &  $W \tau_{\max}$& $T\nu_{\max}$ \\
		\hline
        Spatial  narrowband channels  & $\ll 1$ &  & \\
		Spatial wideband channels & $\geq 1$ &   &\\
        Frequency narrowband channels &  & $\ll 1$ & \\
        Frequency wideband channels &  & $\geq 1$  &\\
        Time-invariant channels &   &      &$\ll 1 $\\
        Time-selective channels &   &      &$\geq 1 $\\
        \hline
	\end{tabular}
\caption{Classification of Wireless Channels on the Basis of Channel and
Signaling Parameters}
\label{table}
\end{table}

It needs to be mentioned that most massive MIMO (millimeter) literatures \cite{MIMO-F1,MIMO-F2, MIMO-F3,  MIMO-F5,MIMO-F6,mmWave1,mmWave2,mmWave3,measure-mmwave,capacity-mmwave, mimo-mmwave}, though assuming $M$ ($W$) to be extremely large, only stick to the spatial narrowband model \eqref{htf-narrow-frequency} and  \eqref{htf-narrow-vector}, while ignoring the spatial wideband phenomenon that actually corresponds to \eqref{httao} and \eqref{steering-narrow}.

Moreover, for  extremely large MIMO system \cite{verylarge-1,verylarge-2,verylarge-3}, the wideband effect or beam squint must be considered, which was unfortunately still ignored in their discussion.
\subsection{Uplink Downlink Reciprocity for Both TDD and FDD}
It has been shown in \cite{angle-delay1,angle-delay2,angle-delay3} that the conductivity and relative permittivity  of  most   materials remain unchanged if the frequency of the electromagnetic wave does not vary much, say less than 1GHz. Hence, the physical AoA of the uplink channel is roughly the same as the AoD of downlink channel in TDD/FDD systems, which is called the \emph{angle reciprocity} \cite{hx-dft}. In fact, angle reciprocity had already been observed in \cite{angle-recip1,angle-recip2} but was not tangibly utilized due to the low angle resolution of the small MIMO. In turn, the path delay $\tau_p$ that is only related to  light speed and the path length is also reciprocal for uplink and downlink channels in FDD systems. Meanwhile, it is obvious that the user's velocity or the scatters' velocity is independent of the carrier frequency.

Hence, although there does not exist direct reciprocity between uplink and downlink channels in FDD systems, the reciprocity can still be built upon the physical parameters like angle, delay, and Doppler shift. Let us use the superscript $(\cdot)^{\text{dl}}$  and $(\cdot)^{\text{ul}}$ to represent the downlink and uplink parameters, respectively. From previous discussion, there are
\begin{equation}\label{psi-dl}
\theta_p^{\text{dl}}=\theta_p^{\text{ul}},\qquad \tau_p^{\text{dl}}=\tau_p^{\text{ul}},\qquad  \frac{\nu_{p}^{\text{dl}}}{f_c^{\text{dl}}}= \frac{\nu_{p}^{\text{ul}}}{f_c^{\text{ul}}}.
\end{equation}

Equalities (\ref{psi-dl}) are also named as the
\emph{angle-delay-Doppler reciprocity} and is important for designing the  massive MIMO systems with advanced array signal processing techniques.
For example, to estimate downlink channel of FDD systems, one only needs to estimate the  channel gain $\beta_p^{\text{dl}}$ if the uplink channel parameters  are known \cite{hx-dft,hx-full,bolei-bs,fandian-60G}. Hence, the  typical overhead restriction of the downlink training in FDD systems can be removed.
In the meantime,  an over-the-air (OTA) test was set up in \cite{NOMP-dongnan} to assess the system performance of the  downlink reconstruction that utilized the parameter reciprocity in practical FDD wireless communication scenarios. The OTA results shown that the  channel reconstructed by the parameter reciprocity is  close to that obtained from linear minimum mean square error (LMMSE) estimator, and higher accuracy can be achieved by  increasing the number of antennas.

In \cite{MIT-reciprocity}, Deepak \emph{et al.}  even demonstrated that in certain ideal scenario,
it is possible to directly predict the $\beta_p^{\text{dl}}$ from  the uplink $\beta_p^{\text{ul}}$ without any additional effort, such that the downlink training can be completely removed.

Angle reciprocity could also facilitate the derivation of the downlink channel covariance matrix (CCM) from the uplink one for FDD systems. In \cite{CCM}, the authors consider the time-invariant spatial and frequency narrowband channel where the uplink and downlink CCMs can be expressed as\footnote{{When $P$ goes to infinite, the sum function changes to integration which corresponds to the continuous incident angles. }}
\begin{align}\label{CMM}
 \bm R_h^{\text{ul}}=\sum_{p=1}^{P} \mathbb E\{|\beta_{p}^{\text{ul}}|^2\} \bm a(\theta_p, f^{\text{ul}}_c,0)\bm a^H(\theta_p,f^{\text{ul}}_c,0),\notag\\
  \bm R_h^{\text{dl}}=\sum_{p=1}^{P} \mathbb E\{|\beta_{p}^{\text{dl}}|^2\} \bm a(\theta_p, f^{\text{dl}}_c,0)\bm a^H(\theta_p,f^{\text{dl}}_c,0).
\end{align}
Clearly, $\bm R_h^{\text{dl}}$ and $\bm R_h^{\text{ul}}$ are not the same but are highly correlated. This property is used to derive $\bm R_h^{\text{dl}}$ from $\bm R_h^{\text{ul}}$ in an easier way \cite{CCM}. Hence, the conventional training  spent on the downlink covariance estimation and feedback can be well saved.

\subsection{Low-Rank CCM Property of Massive MIMO}
With array signal processing, one can well link the angle domain property with the rank  of the CCM. Considering  a finite scattering environment with $P$ AoAs, $\theta_{p}^{\text{ul}}\in [\theta_{\min}^{\text{ul}},\theta_{\max}^{\text{ul}}]$, $p=1,2,\cdots,P$, it is proved in \cite{lowrank2} that the rank of  $\bm R_h^{\text{ul}}$ satisfies
\begin{align}\label{CMM-low}
\!\!\frac{ { \text {rank}(\bm R_{h}^{\text{ul}})}} {M} \leq |\sin(\theta_{\max}^{\text{ul}})\!-\sin(\theta_{\min}^{\text{ul}})|\frac{d}{\lambda_c}, \quad \!\!\!\text{as} \!\!\!\quad\! M\rightarrow \infty,
\end{align}
which tells that  the rank of $\bm R_h^{\text{ul}}$ would be proportional to the angular spread (AS) $|\sin(\theta_{\max}^{\text{ul}})\!-\sin(\theta_{\min}^{\text{ul}})|$ of all multi-paths (similar property holds for downlink CCM). Obviously,  $\bm R_h^{\text{ul}}$  would be of low rank when
AS is small\footnote{
Here, we  assume: (i) The antenna spacings and the end-to-end distance do not satisfy the \emph{Rayleigh distance criterion} \cite{Res-1}; (ii) The mmWave massive MIMO channel is not under the ``best-case rain rate" \cite{Res-2}. Otherwise, mmWave massive MIMO channels may also be of high rank.}. The low-rank property of CCM has been leveraged by many existing works to handle the pilot contamination and to solve the downlink channel estimation issue. The required narrow AS can be observed in the following scenarios:
(i) BS equipped with a large number of antennas is always
elevated at a very high altitude, such that there are few surrounding scatterers \cite{MIMO-sparsity1,MIMO-sparsity2,MIMO-sparsity3};
(ii) When massive MIMO system is employed at the mmWave band, the high path
loss leads to limited number of the incoming  paths  \cite{mmWave1,mmWave2,mmWave3,mmwave-survey}.
\section{Parameters Estimation}
Channel estimation has long been deemed as a bottleneck for massive MIMO systems due to a large number of unknowns  to be estimated, especially for the downlink case. For a rich scattering environment, the channel covariance is full rank and there is no way to save the channel estimation effort. In this case, the conventional least square (LS) or LMMSE based channel estimation has to be applied for massive MIMO \cite{MIMO-sync,MIMO-user,MIMO-F1,MIMO-F5}.

Nevertheless, a major concern is that the massive MIMO systems are normally launched in high attitude or in mmWave frequency band such that low-rank CCM or the channel sparsity regularly holds \cite{MIMO-sparsity1,MIMO-sparsity2,MIMO-sparsity3}. With sparse assumption, many channel estimation algorithms that aim to reduce the training overhead have been proposed from communication theory viewpoint, such as virtual MIMO channel representation (VCR) model  \cite{virtual-channel} and compressive sensing (CS) methods \cite{sayeed-4D, CS-method}. A detailed overview of sparse channel estimation methods in massive MIMO can be found in \cite{Liye-survey,hx-overview}  and will not be discussed here. The basic idea is to take discrete Fourier transform (DFT) matrix as the dictionary and apply the conventional CS algorithm to recover the channel.  However, such  idea is equivalent to assuming that the $P$ AoAs of the signal paths reside  exactly on fixed grids $\{0, \frac{2\pi}{M},2\frac{2\pi}{M},\ldots,(M-1)\frac{2\pi}{M}\}$,
which could only provide an approximate channel model. \footnote{{If $P$ is extremely large or the incident angles are in the continuous domain, it is necessary to approximate all incident angles with the equivalent finite discrete incident angles.}}
This is known as  \emph{grid mismatch} \cite{grid-mismatch,grid-bias}  and would cause power leakage problem. Consequently, the performance of the channel estimation with VCR would reach an error floor in  high  SNR regions.

Alternatively, instead of estimating the channel $\mathbf{h}$ (communication theory viewpoint), one can also choose to estimate the parameters inside the channel, i.e., $\theta_p$, $\tau_p$, $\nu_{p}$, $\beta_p$, as has been tried in conventional smart antenna. However, since the number of antennas $M$ in conventional MIMO is small and the number of paths $P$ is large in microwave band, it is impossible to accurately extract those parameters.

Nevertheless, with massive antenna array and sparse assumption (especially the mmWave scenario),  it is possible to apply array signal processing techniques to extract precise AoA/AoD parameters from the channel. Moreover, It should be emphasized that only half-wavelength arrays can estimate AoA without ambiguity, while for those massive MIMO systems with larger inter-antenna distances \cite{MIMO1,MIMO2,MIMO3,MIMO-sync,MIMO-user,MIMO-F1},
it is impossible to apply AoA/AoD based signal processing.
Fortunately, because of the size limitation, the majority of massive arrays are equipped with  half-wavelength  in practice.
Similarly,  with broad communication bandwidth and the sparsity assumption, it is also possible to apply array signal processing techniques to extract precise delay and Doppler parameters \cite{mmWave1,mmWave2,mmWave3,measure-mmwave,mmwave-survey}.
\subsection{Blind Parameters Estimation }
The blind AoA estimation is a classical problem in the area of array signal processing, and there are many well-known approaches  based on the eigen-value decomposition (EVD) of the signal covariance matrix (not CCM).
For example, the parametric algorithms MUSIC \cite{music} and ESPRIT \cite{esprit} have been already demonstrated their superior resolution   compared to the non-parametric counterparts, say the DFT-based ones.

With massive MIMO configuration, \cite{1D-esprit1}  compared the Root-MUSIC with the ESPRIT algorithms at frequency 30~GHz with  a 1D ULA of 72 elements. It was shown that the root-MUSIC algorithm outperformed the ESPRIT under good radio channel conditions. Meanwhile, Yong \emph{et al.} \cite{1D-esprit2} proposed a power profile based AoA estimation (PROBE) algorithm for 1D Lens-embedded mmWave MIMO systems. In terms of computational complexity, the blind PROBE algorithm has an  advantage over the MUSIC and ESPRIT algorithms.
Due to the size restriction, the antenna arrays of massive MIMO systems were expected to be implemented in more than one dimension \cite{2D-array,2D-esprit1,2D-esprit2 }. The work \cite {2D-esprit2} proposed an ESPRIT-based approach for 2D localization of multiple sources employing very large uniform rectangular arrays (URAs). To estimate  nominal AoAs, which were coupled in the array steering vector, the array has to be divided into at least three subarrays to decouple the 2D nominal AoAs.
However, such a 2D model is still used to estimate azimuth AoAs only, while is not adequate for 3D channel estimation or the so-called full-dimension MIMO (FD-MIMO) \cite{fd-mimo}.
The 3D models took into account azimuth as well as the elevation of signal propagation. The works \cite{3D-esprit1,3D-esprit2} then applied the ESPRIT algorithm to estimate the 3D channel parameters, such as  the azimuth AoAs, the elevation AoAs, and the  azimuth  AoDs in the mmWave massive MIMO systems.

On the other side, the delay factor can also be estimated via array signal processing techniques, especially for frequency-wideband  systems.
Compared with the methods only estimating AoAs \cite{1D-esprit1,1D-esprit2,2D-array,2D-esprit1,2D-esprit2,3D-esprit1,3D-esprit2}, joint AoA and delay estimation methods have shown significant superiority in terms of the accuracy of the reconstructed channel, since both the spatial and temporal diversity of the multipath channels were exploited \cite {doa-delay1,doa-delay2,doa-delay3}.
Specifically, with the unitary ESPRIT, \cite{doa-delay3} investigated  joint/separated angle/delay estimation methodologies in 3D massive MIMO millimeter wave systems.
In addition, \cite{doa-doppler} utilized the multi-dimensional
unitary ESPRIT  to extract the azimuth and elevation  parameters as well as the Doppler frequencies from the noisy channel, where URA was adopted at both the transmitter  and receiver of the  mmWave MIMO systems.

The blind  algorithms  highly rely on the statistical characteristics of the received signals and thus  have high spectral efficiency \cite{blind1}. However, these algorithms often require a large number of received signals, which are restricted  to slow time-varying channels and also encounter high complexity \cite{pilot-design}.
\subsection{Training Based Method}
Different from conventional array signal processing, wireless communications offers  cooperate terminals and it is possible to use training sequence for more efficient parameter estimation \cite{pilot-design2}, \cite{pilot-design3}, while leaving blind way as the supplement for parameters tracking.
\subsubsection{CCM Based Method}
Consider a finite scattering environment for massive MIMO systems and  assume that the AS of each user is restricted within  a narrow region.
Based on  \eqref{CMM-low}, we define
\begin{align}\label{rank}
 r_k \triangleq \text{Rank} (\bm R_k) \ll M,
\end{align}
where $\bm R_k$ is the CCM of the $k$th user with size $M\times M$.
Then, the channel of the $k$th user can be  expanded by $r_k$ dominant eigenvectors that correspond  to $r_k$ nonzero eigenvalues, which would reduce the  channel dimensions from $M$ to $r_k$.
One could mathematically demonstrate the low-rank property of CCM as \cite{JSDM}
\begin{align}\label{svd}
\bm R_k =\bm U_k \bm \Lambda_k \bm U_k^H,
\end{align}
where $\bm \Lambda_k $ is the nonzero eigenvalue matrix of size $r_k \times r_k$ and $\bm U_k$ is the subspace eigen-matrix of size $M\times r_k$.

It indicates that the CCMs of any two users with non-overlapped AS are asymptotically orthogonal to each other\cite{hx-overview}, i.e.,
\begin{align}\label{Uk}
\bm U_k^H\bm U_l \rightarrow \bm 0, \quad \text{for}\quad \mathcal A_k \cap \mathcal A_l= \varnothing, \text{as} \quad M\rightarrow \infty,
\end{align}
where $\mathcal A_k$ and $\mathcal A_l$  are the continuous AS intervals for the $k$th and the $l$th users, respectively.
Hence, the pilot contamination could be removed for
these users with non-overlapped AS even if they employ
the same training sequence. Based on \eqref{Uk},  \cite{JSDM}  proposed a downlink joint
spatial division multiplexing (JSDM) scheme, where a classical multiuser precoder was adopted to restrict each user's
beamforming vectors within the orthogonal complement of
the channel subspaces of the others. Meanwhile, \cite{lowrank2} directly
applied uplink channel training via the minimum mean square
error (MMSE) estimator and proved that channels with
non-overlapped AS can be estimated free of interference.

By leveraging the low-rank property of CCMs and reducing
the effective dimensions of channels, downlink pilot contamination
as well as downlink training and feedback overhead can be significantly reduced \cite{hx-overview}.
However, the acquisition of CCM is a difficult task for multi-user massive MIMO
systems because each user's high-dimensional downlink CCM  has to be separately estimated and then fed back to BS. Furthermore, the accompanied computational complexity involved in the EVD of high-dimensional CCMs for multiple users is hardly affordable.
\subsubsection{DFT Based Algorithms}
The spatial and frequency narrowband channel \eqref{htf-narrow-vector}  naturally connects to Fourier transform in that when $M$ goes to infinity, the DFT of $\mathbf{h}(t)$ in  \eqref{htf-narrow-vector}  would only presents $P$ non-zero peaks at the corresponding  AoAs \cite{hx-dft}. Hence  the DFT-based representation of the massive MIMO channel with $P$ non-zero supports are usually used as an approximation of $\mathbf{h}(t)$, i.e., the so called VCR \cite{virtual-channel}:
\begin{align}\label{hf-narrow-dft-rotation}
 {\bm h(t)}\thickapprox  \bm W_M^H  \bm b,
\end{align}
where $\bm W_M \in \mathbb C^{M\times M}$ is the normalized DFT matrix whose $(m,n)$th entry is  $[\bm W_M]_{m,n}=e^{-j\frac{2\pi}{M}mn}/\sqrt{M}$, $0\leq m,n\leq M-1$, and $ \bm b\in \mathbb C^{M\times 1}$ is assumed to have $P$ non-zero entries with the position related to $\theta_p$.  The corresponding method is named as \emph{beamspace} channel estimation \cite{virtual-channel,beamspace-sayeed1,beamspace-sayeed2}.

\begin{figure}[t]
\centering
\vspace{-0.5cm}
\subfigure[An ideal power spectrum with no leakage, $\theta=30^{\circ}$.]{
\includegraphics[width=0.45\textwidth]{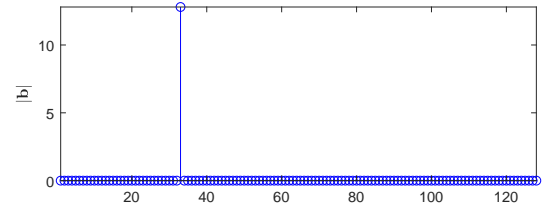}
\label{dft1}
}
\subfigure[An leakage-existing power spectrum, $\theta=30.5^{\circ}$.]{
\includegraphics[width=0.45\textwidth]{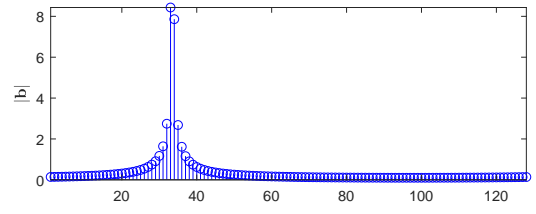}
\label{dft2}
}
\caption{Illustration of two power spectrums   with $M =128$ and $P=1$. }
     \label{dft-power}
\end{figure}

However, DFT could only yield the AoA estimation with the resolution of $1/M$ \cite{M1} and is a non-parametric AoA estimation method. Consequently, due to the mismatch between DFT basis and real $\theta_p$,  the power of the beamspace subchannels would leak to their neighbors, and hence the true number of non-zero supports, i.e., the non-zero entries in $\bm b$ is much larger than $P$, as shown in Fig. \ref{dft2}. The resultant phenomenon is also called grid mismatch or \emph{power leakage effect} \cite{hx-overview}, which leads to significant channel estimation error or the increased training overhead in order to estimate all the non-zero supports.

The authors of \cite{linhan-dft} then proposed to add zero at the end of $\mathbf{h}(t)$ before the DFT operation, such that the angle  domain is sampled in a much denser way. In this case, a better angle estimation can be obtained compared to beamspace method.  On the other side, the authors \cite{hx-dft}  proposed to use a spatial rotation to mitigate the power leakage effect by rotating the channel vector and concentrating most power in much fewer grids.
Compared to beamspace model,  \cite{linhan-dft} and \cite{hx-dft} targeted at the real AoA estimation and was named as \emph{anglespace} modeling or the \emph{modified VCR}.
With accurate AoA estimation,  \cite{hx-dft} further applied the \emph{angle reciprocity} to simplify the  downlink channel estimation for both TDD and FDD systems.
The work \cite {fandian-60G} then extended anglespace method to indoor mmWave  massive communications with URA.

For frequency wideband systems, the author  \cite{yangCY} utilized the 2D DFT-based algorithms to capture the hidden sparsity of the channel and estimate the on-grid AoAs and delays.
For dual-wideband systems,  \cite{bolei-bs} utilized the 2D DFT-based algorithm together with 2D rotation to estimate the AoAs and delays of the multi-paths.
With accurate AoA and delay estimation, \cite{bolei-bs} further applied the
\emph{angle-delay reciprocity} to simplify the  downlink channel estimation in  FDD systems.
\subsubsection{Compressive Sensing Based Algorithms}
CS provides a sparse signal recovery method in the mmWave massive MIMO systems and is a relatively new area of signal processing. See \cite{CS-overview1,CS-overview2,CS-overview3} for the overviews of the fundamental theory in CS.

Based on spatial- and frequency-narrowband  steering vector $\bm a (\theta_p,0,0)$, the channel vector in \eqref{htf-narrow-vector} can be  represented by the sparse format \cite{CS-eldar-book}:
\begin{align}\label{hls-cs-on}
\bm h (t)= \bm A(\bar {\bm \theta}) \bm x+ \bm n,
\end{align}
where the $M\times L$ measurement matrix $\bm A(\bar {\bm \theta})$ is defined as $\bm A(\bar {\bm \theta}) \triangleq[\bm a(\bar\theta_1,0,0),\cdots,\bm a(\bar \theta_L,0,0)]$  and $\bar {\bm \theta}$ is defined as $\bar {\bm \theta}\triangleq[\bar\theta_1,\bar\theta_2,\cdots,\bar\theta_L]$ with $\bar\theta_i= {-\pi}+\frac{2i\pi}{L},  i\in\{1,2,\cdots,L\}$  dividing the continuous angle space uniformly. Moreover, $\bm x  \in \mathbb C^{L\times 1}$ has the sparsity level of
$P$, i.e., $\bm x$  has only $P \leq M$ non-zero elements. Similar to beamspace method \cite{virtual-channel,beamspace-sayeed1,beamspace-sayeed2}, the CS model (\ref{hls-cs-on}) assumes that the true  channel parameters are exactly aligned with the grids but with a finer grid interval when $L\geq M$. Then, the AoA estimation problem is formulated as:
\begin{align}\label{lo}
\hat { {\bm \theta}}= \min_{\bm{x}} \| \bm x \|_{0}, \qquad
             \text {s. t. }  \|\bm h(t) - \bm A(\bar {\bm \theta}) {\bm x}\|_2\le \xi,
\end{align}
where $\xi$ is an error tolerance parameter that is related to the noise statistics.
Then, the CS-based sparse channel estimation boils down to angle estimation and path gain estimation subproblems, which can be solved sequentially with reduced problem complexity \cite{GS16CS}.
Note that the on-grid CS model considers the $\bar {\bm \theta}$ as the known parameter whose accuracy depends on  $L$. It only provides the true AoA estimation
when $L$ goes to infinity but the accompanied complexity is unaffordable. To deal with this problem,  \cite{Fangjun}  treated $\bar {\bm \theta}$ in \eqref{hls-cs-on} as the unknown continuous parameters that can take arbitrary values and  covert the objective to  not only estimating the sparse signal, but
also optimizing/refining the  parameters $\bar {\bm \theta}$. Then, the problem becomes a  sparse signal recovery problem with an off-grid parametric dictionary. i.e.,
\begin{align}\label{lo-off}
\hat { {\bm \theta}}_{\text{off}}= \min_{\bm{x}, {\bm \theta}} \| \bm x \|_{0}, \qquad
             \text {s. t. }  \|\bm h(t) - \bm A ({\bm \theta}) {\bm x}\|_2\le \xi,
\end{align}
where $\bm \theta$ is the $P\times 1$ unknown angle vector, and $\bm A( {\bm \theta})$ is the $M\times P$ off-grid parametric
containing $P$ columns of steering vectors.

For either the on-grid or  off-grid AoA estimation, many CS sparse signal recovery algorithms can be utilized and  are mainly
divided into the following four categories:
\paragraph{Convex Relaxation Algorithms}
A famous convex relaxation for the sparse recovery algorithm is the  least absolute shrinkage and
selection operator (LASSO).
Donoho \cite{CS-overview1} and Candes \cite{Candes} showed  that one avenue for translating \eqref{lo} into a tractable problem is to replace $\|\cdot\|_0$ with its convex approximation $\|\cdot\|_1$.
The LASSO estimate of \eqref{hls-cs-on} is formulated as a Lagrangian relaxation of a quadratic program, which is given by \cite{lasso}
\begin{align}\label{LASSO}
\hat {\bm x}= \lambda \sigma \min_{\bm{x}} \| \bm x \|_{1}+ \frac{1}{2} \| \bm h(t) - \bm A(\bar{\bm \theta}) {\bm x}\|_2,
\end{align}
where $\sigma$ is the standard derivation of the noise, and $\lambda >0 $ is the regularization parameter.
Sayeed \emph{et al.}  \cite{lasso-ongrid} then utilized the  LASSO-based algorithm for the sparse virtual  channel estimation,
which exploited  the inherent sparsity of mmWave channels in the angle and delay
domains.
Eltayeb \emph{et al.} \cite{lasso-heath} also developed a LASSO based array diagnosis
techniques for the mmWave systems with large antenna arrays which jointly estimated the  AoAs/AoDs and the phase-shifts.
\paragraph{Greedy Algorithms}
Greedy iterative algorithms  select columns of $\bm A( \bar {\bm \theta})$ according to their inner product correlation with the
measurements $\bm h (t)$ in a greedy iterative manner \cite{CS-eldar-book}.
At each iteration, the sparse vector is updated from
\begin{align}\label{omp-x}
\min_{\bm x_\Lambda} \|  \bm h(t) - \bm A_{\Lambda}(\bar{\bm \theta}) {\bm x_\Lambda}  \|^2,
\end{align}
where $\bm x_{\Lambda}=\bm A_{\Lambda}(\bar {\bm \theta})^{\dagger}\bm h (t) $, and $\Lambda$ denotes the current estimated support set. Thus, in each iteration, the residual $\bm h (t) - \bm A_{\Lambda}(\bar {\bm \theta}) {\bm x_\Lambda}$ is orthogonal to the columns of $\bm A(\bar{\bm \theta})$  that are included in the current estimated support \cite{CS-eldar-book}.
Lau \emph{et al.} \cite{on-omp}  considered the multi-user massive MIMO systems and deployed an orthogonal matching pursuit (OMP)-based technique to reduce
the training as well as the feedback overhead for AoA estimation.
However, the conditions on OMP estimation algorithms are more restrictive than the restricted
isometry condition (RIC) \cite{RIC}, which can easily lead to weak parameter estimation.
Gui \emph{et al.} \cite{on-cosamp} then utilized a robust compressive sampling matching pursuit (CoSaMP)-based algorithm to exploit the block-structure sparsity in angular domain in order to  further improve channel estimation of massive MIMO systems.
However,  \cite{on-omp,on-cosamp} still  face the grid mismatch problem.
Then, Wang \emph{et al. } \cite{Mingjin} proposed
a shift-invariant block-sparsity  based channel estimation algorithm
which jointly computed the off-grid angles, the off-grid delays, and the complex gains of the wideband mmWave massive MIMO channels.
\paragraph{Atomic Norm Algorithms}
In order to circumvent   grid mismatch, \cite{AND} proposed an algorithm named  atomic norm denoising, expressed as
\begin{equation}\label{eq:pri}
\hat {\bm x} =\argmin\limits_{{\bm x}}\ {\frac{1}{2}} \| \bm A(\bar {\bm \theta}){\bm x} - {\bm h(t)}\|_{2}^{2} + \gamma \Vert {\bm x}\Vert_{\cal A},
\end{equation}
where $\gamma$ presents the regularization parameter and $\Vert\cdot\Vert_{\cal A}$ denotes the  atomic norm defined in \cite{AND}. Bhaskar \emph{et al.} \cite{AND} proved that \eqref{eq:pri} can be converted into a semi-definite programming that can be computed by off-the-shelf solvers such as SeDuMi and SDPT3 in CVX toolbox.

Atomic norm denoising can be applied to estimate the channel parameters with super-resolution \cite{AND} and was firstly applied for super-resolution channel estimation in mmWave massive MIMO systems by Wang \emph{et al.} in \cite{ICC17ANM}. To avoid the enlarged problem size due to the vectorization-based atomic norm minimization in the 2D-angle scenarios, Tian \emph{et al.} \cite{ICASSP17DANM} proposed a decoupled atomic norm minimization technique and its theoretical results were provided in \cite{SP19DANM}.
Moreover,  Tsai \emph{et al.} \cite{chesti} proposed an atomic-norm-denoising-based method for mmWave FD-MIMO channel estimation.
Tirkkonen \emph{et al.}  \cite{precod} formulated  mmWave MIMO channel estimation as atomic norm denoising problem and then designed the multi-user precoder.
Besides, Chu \emph{et al.} \cite{atomic-1} utilized the atomic norm minimization to manifest the channel sparsity in the continuous azimuth AoAs and AoDs, which also provided a super-resolution channel estimators for mmWave MIMO systems.

However, the deficiency is that atomic norm denoising algorithm becomes slow when solving large scale problems. In order to provide a faster method for the semi-definite program, \cite{AND,chesti} adopted the alternating direction method of multipliers (ADMM) to accelerate such process.

\paragraph{Bayesian  Algorithms}
The sparse Bayesian learning (SBL)  principle could infer the sparse unknown signal from the Bayesian viewpoint by considering the sparse priori  \cite{bayesian1}.
A typical SBL model is  given by \cite{bayesian1,bayesian2}:
\begin{equation}\label{bayes}
p(\bm h|\bm x,\zeta,\bm A(\bar {\bm \theta}))= \mathcal {CN} (\bm A(\bar{\bm \theta})\bm x,\zeta^{-1}\bm I_M),
\end{equation}
where $\zeta \triangleq \sigma^{-2}$  denotes the noise precision. In general, $\zeta$ is assumed to  follow  Gamma prior distribution $p(\zeta)=\Gamma (v;\chi,\upsilon)$, where $\Gamma(\cdot)$ is the Gamma function, $\chi$ is the shape parameter, and $\upsilon$ is the rate parameter.

Matteo \emph{et al.} \cite{bayesian-on-doa-narrow} first utilized the SBL method to estimate the AoA from the narrowband MIMO channels.
Meanwhile, Zhang \emph{et al.} \cite{bayesian-on-doa-wide} proposed  a sparse Bayesian-based estimation algorithm for frequency wideband communications, which took advantage of the band occupation information by leveraging the cluster property of the Dirichlet process. However, \cite{bayesian-on-doa-narrow,bayesian-on-doa-wide} are still on-grid sparse Bayesian algorithms.  Das and Terrence  \cite{bayesian-off1} then proposed an off-grid version of SBL based relevance vector machine algorithm (SBLRVM) to estimate the AoAs of multiple narrowband signals. Jian \emph{et al.} \cite{bayesian-mengnan} utilized the off-grid SBL method to obtain super-resolution angle and delay for the accurate channel reconstruction in dual-wideband mmWave massive MIMO systems.
The off-grid Bayesian algorithms have  superior performance due to their use of data-adaptive priors, which is suitable  to extract the unknown parameters i.e., AoAs and  delays from the  structured array.
\vspace{-2mm}
\subsection{Prospects of Parameters Estimation}
The core idea of parametric massive MIMO channel modeling is to estimate the physical channel parameters using various effective manners.
However, it should be emphasized that though the parametric channel estimation algorithms can greatly reduce the computational complexity brought by the massive MIMO systems, they are very sensitive to the number of multi-path and the accuracy of estimated parameters.
There will be a distinct angle error when the estimated path number is wrong, especially in the discrete sparse scenes with finite incident paths. This is also a critical  problem in traditional array signal processing techniques. The commonly accepted algorithms for path number estimation are minimum description length (MDL) and Akaike information criterion (AIC) \cite{AIC},
while MLD and AIC algorithms both have upper precision limit.
At present, how to design a low-complexity as well as high-resolution algorithm is still a challenging work, and we hope researchers can  conduct in-depth work in this area.
\vspace{-2mm}
\section{Synchronization}
Apart from the channel estimation, another challenge of massive MIMO is the synchronization issue \cite{sync1,sync2}.
Particularly for  mmWave massive MIMO,  the Doppler spread  is orders-of-magnitude larger than that of classical narrowband wireless channels, which may deteriorate the system performance \cite{sync-GaoXQ}.

Consider wideband  transmission employing OFDM modulation. With perfect time and frequency synchronization in the space domain, the length of the cyclic prefix (CP) is usually set to be slightly larger than the delay span to mitigate time  dispersion, while  the length of the OFDM symbol is usually set to be inversely proportional to the Doppler spread to mitigate frequency dispersion \cite{sync3,sync4}.
Embracing  the beam squint effect, the authors of \cite{bolei-bs} showed that extra CP length is required to overcome the time delays across the large array aperture.
As a result, the overhead caused by CP will be much larger to deal with the same delay spread for wideband massive MIMO system.

You \emph{et al.} \cite{sync-GaoXQ} proposed per-beam synchronization (PBS) for mmWave massive MIMO-OFDM transmission, where both delay and Doppler frequency spreads of the wideband MIMO channels  were reduced  approximately by a factor of the number of antennas.
In the angle domain, Zhang \emph{et al.}  \cite{sync-ZhangWL} designed  a frequency synchronization scheme for multi-user OFDM uplink with a massive ULA at BS.  By exploiting the angle information of users, the carrier frequency offset (CFO) was estimated for each user individually through a joint spatial-frequency alignment procedure.
Moreover, the authors in \cite{sync-ZhangWL2} and \cite{sync-ZhangWL3} addressed the joint estimation issue of Doppler shifts and CFO in high-mobility downlink massive MIMO systems, where a high-resolution beamforming network was designed to separate different Doppler  and CFOs in the angle domain.
\vspace{-3mm}
\section{Beamforming and Precoding}
The precoding techniques for massive MIMO systems with both sub-6 GHz and mmWave frequency band have  been well studied in \cite{Liye-survey,Heath-survey,precoding-Jin1,precoding-Jin2,precoding-Jin3,precoding-Jin4}. In this section, we will not go through those works that have already been described from communications viewpoints but
rather to introduce those designed from the array signal processing viewpoint. As mentioned in Section II, the array signal processing based design is mainly
applicable  for large arrays and sparse environments.
In fact many beamforming designs for mmWave originate from array signal processing techniques.
\subsection{Hybrid Precoding}
In conventional MIMO systems, each antenna is driven by its own dedicated radio frequency (RF) chain and thus the full digital beamformer (DB) can be applied \cite{beam-survey}. However, DB is not suitable for practical massive MIMO systems since the huge number of RF chains brings high hardware cost, high system complexity, and large power consumption. The hybrid analog and digital beamforming architecture is then proposed, where $M$ antennas share the same $M_{\text{RF}}\ll M$ RF chains \cite{hybrid1}, \cite{hybrid2}. The analog beamforming matrix is usually realized by a low-cost phase shifter or switching network that imposes a constant modulus constraint on each element.

For conciseness, we illustrate the beamforming design with  a single user only, as did in \cite{Heath-survey}. Under spatial and frequency narrowband environment, the downlink transmission can be expressed as
\begin{align}\label{hybrid}
\tilde{\bm y}=\bm h^{H}\bm F_{\text{RF}} \bm f_{\text{BB}} s +\bm n,
\end{align}
where $ \bm h=\sum_{p=1}^{P}  \beta_{p}^{\text{dl}} \bm a(\theta^{\text{dl}}_{p},0,0)$ is the downlink channel,  $\bm F_{\text{RF}}$ is the  $M\times M_{\text{RF}}$ analog precoding matrix with constant modulus entries, while $\bm f_{\text{BB}} $ is the $M_{\text{RF}}\times 1$ digital precoding vector.
The objective is to design the hybrid analog and digital beamforming  that can  maximize the overall system spectral efficiency, i.e.,
\begin{align}
\min_{\bm F_{\text{RF}}, \bm f_{\text{BB}}}\  &|\bm h-\bm F_{\text{RF}} \bm f_{\text{BB}}|^2\label{eq:Gao2}\\
\textup{s.t.}\ \  &|[\bm F_{\text{RF}}]_i|=\frac{1}{\sqrt{M}},i=1,2,...,M.\notag
\end{align}
\begin{figure}[t]
\centering
\vspace{-0.6cm}
\subfigure[Non-orthogonal angle space (anglespace) beamforming.]{
\includegraphics[width=0.45\textwidth]{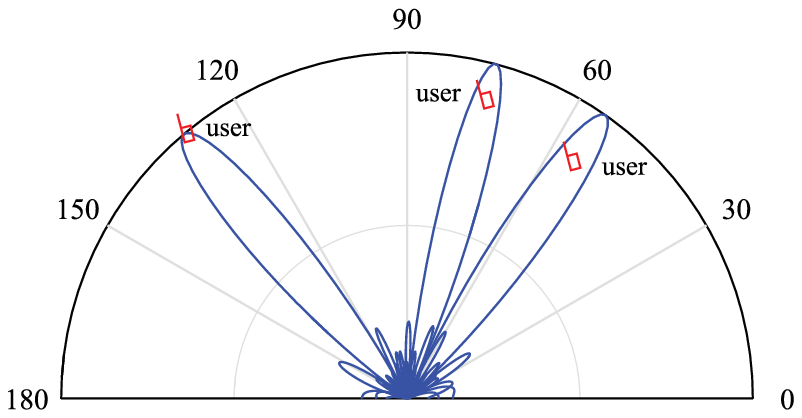}
\label{anglespace}
}
\subfigure[Orthogonal angle space (beamspace) beamforming.]{
\includegraphics[width=0.45\textwidth]{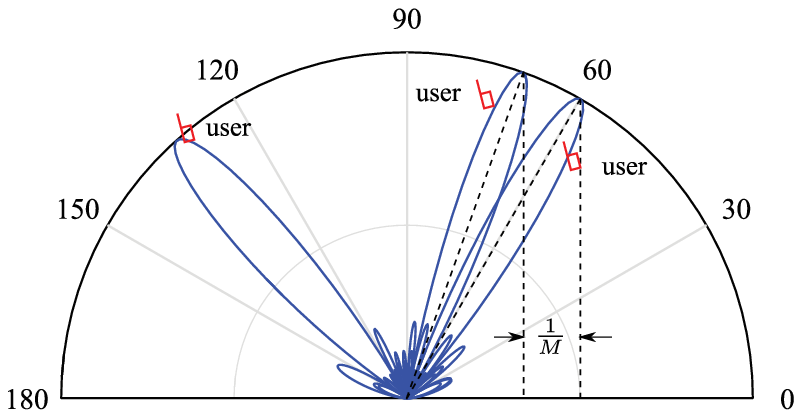}
\label{beamspace}
}
\caption{Illustration of two different beamforming methods. }
     \label{fig:usergrouping}
\end{figure}
The optimization (\ref{eq:Gao2}) can be well addressed from array signal processing viewpoint in that when columns of $\bm F_{\text{RF}}$ contain the steering vector $\bm {a}(\theta_p^{\text{dl}},0,0)$ and elements of $\bm f_{BB}$ match $\beta_p$, then  the objective function  could be minimized. If $M_{RF}<P$, then additional effort should be applied to minimize the objective function, say the beamforming design \cite{Gershman1}. Matching the columns in $\bm F_{\text{RF}}$ with the true steering vector $\bm {a}(\theta_p^{\text{dl}},0,0)$ is named as \emph{anglespace} beamforming, and the corresponding beams are shown in Fig. \ref{anglespace}, \cite{anglespace-jianwei,linhan-dft,hybrid3}. Alternatively, if  $\bm F_{\text{RF}}$  only supports
$M$ fixed directions, then the best beamforming vectors should be selected from a predetermined codebook, i.e., \emph{beamspace} beamforming, \cite{beamspace-sayeed1,beamspace-sayeed2,beamspace-Heath,beamspace-Xiao}. Obviously, the beamspace precoding
can be achieved with low complexity but can only be viewed as ``angle on the grid'' approach and will suffer from power leakage \cite{hx-overview,linhan-dft}, as shown in Fig. \ref{beamspace}.   Once the analog beamforming matrix $\bm F_{\text{RF}}$ is obtained, the digital precoder $\bm f_{BB}$  can be derived  following minimum mean square error (MMSE) criterion  \cite{hybrid-mmse,hybrid4,linhan-dft} or the zero-forcing (ZF) criterion \cite{anglespace-jianwei}.

It has been shown in \cite{Heath-survey} that the hybrid analog and digital beamforming design for multiuser massive MIMO system can reduce the number of RF chains and achieve near-optimal performance as that of fully-digital beamforming structure when carefully designed.
Moreover, there are many other solutions that design the hybrid structure with AoAs/AoDs.  Details can be found in \cite{Heath-survey} and will not be re-stated here.
\vspace{-3mm}
\subsection{Interference Control}
From communications viewpoint, the user interference is due to the non-zero inner product between the channel vector and the beamforming vector, while the cancellation of the interference to a specific user is achieved by adjusting the weight on different antennas. From array signal processing viewpoint, such interference is due to the side lobes of the beamforming vector and leaks towards the undesired user, while the interference cancellation is equivalent to formulating a physical null over the beam pattern towards the AoAs of the undesired users.
To be specific, the lobe corresponding to the
direction of the maximum gain is referred to as the \emph{main lobe} while lobes with much smaller
gain are the \emph{side lobes}. 
Usually, there exists a large ripple in main/side lobes  of hybrid  beamforming structure,
which would cause   interferences between different users.
Suppressing the ripple is  beneficial to achieve a uniform beamforming performance, which can guarantee user fairness to some extent.

\begin{figure}[t]
  \centering
  \vspace{-0.5cm}
  \includegraphics[width=80mm]{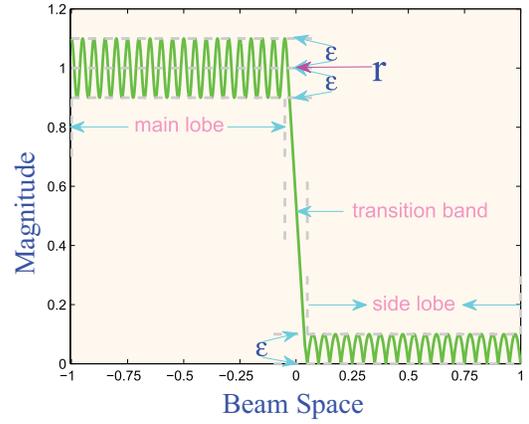}
  \caption{A design example, maximize the magnitude response in the main
lobe, or equivalently $r$, while suppressing the ripple into a given level $\varepsilon$. }
  \label{ripple}
\end{figure}

Ideally, one would expect the ripple of each beam to be as small as possible while the main lobe power as large as possible. However, these two goals generally conflict with each other unless the number of antennas $M$ goes to infinity. To reach a reasonable solution, a small ripple in both the main lobe and the side lobes  are accepted but the ripple must be limited to a small positive real number $\varepsilon$, as shown in Fig. \ref{ripple}.  The design of the training beamforming vector can be obtained from the following optimization problem \cite{sidelobe-equation1,sidelobe-equation2,sidelobe-HuangYM}:
\begin{align}
 &\max_{\bm{f}_{s,b}^{q},r_{s,b}^{q}} r_{s,b}^{q}\notag\\
 {\text{s. t.} } \quad   &|\bm a(\theta_p,0,0)\bm f_{s,b}^q- r_{s,b}^{q} |  \leq \varepsilon, \quad (\theta_p \in I_{s,b}^q)  \notag\\
 &|\bm a(\theta_p,0,0)\bm f_{s,b}^q |  \leq \varepsilon, \quad (\theta_p \not\in I_{s,b}^q),
\end{align}
where $r_{s,b}^{q}$ is the main lobe power, $\bm f_{s,b}^q$ is the training beamforming vector, and $I_{s,b}^q $ is the beam interval.

Alkhateeb \emph{et al.} \cite{beamspace-Heath}
have realized the importance of controlling the ripple. However, the designed
interference control method from LS principle in \cite{beamspace-Heath} cannot well suppress the ripple and thus cause a
large performance loss on non-quantized points in the beamspace
domain. Raghavan  \emph{et al.} \cite{sidelobe-1} proposed a maximin optimization to
achieve a well-shaped training beam, which essentially looked
for a flat array gain response in the coverage area of the
designed beam and thus can realize  a small ripple in the main
lobe as well as side lobes.
To reduce the user interference and increase the beamforming performance,  Zhang {\emph{et al.} \cite{sidelobe-HuangYM} proposed a training codebook design by shaping the geometric pattern of each beam and formulating the beam design as an optimization problem with both the ripple and the transition band being taken into consideration.
\vspace{-2mm}
\section{Orthogonal Time and Frequency Space of Massive MIMO}
In addition to exploiting angle, delay and Doppler separatively, a new modulation named orthogonal time frequency space (OTFS) considers all of these three parameters in  massive MIMO systems.
Leveraging the basis expansion model (BEM) for the channel \cite{BEM}, OTFS first  transforms the time-varying multipath channel into a two-dimensional time-independent  channel in the delay-Doppler domain \cite{otfs-original,otfs-original2}.
The OTFS  rearrange the data sequence of length $N_e N_o$ into a 2D data block with the size of $N_e\times N_o$, where $N_e$ and $N_o$ are the number of units along the delay dimension and Doppler dimension. This block data is transmitted with total duration $N_e\triangle T$ seconds and total bandwidth $N_o \triangle f$ Hz, e.g, a sampling of the time and frequency axes at intervals $\triangle T$ and $\triangle f$, respectively.
OTFS is different from previous work in that it multiplexes data in
the delay-Doppler domain, and each transmitted symbol experiences a near-constant channel gain
even in channels with high Doppler fading \cite{otfs-doppler}, massive MIMO \cite{otfs-MIMO}, or at high frequencies such as millimeter waves \cite{otfs-mmwave}.
The relatively constant channel gain over all symbol transmissions obtained in OTFS massive MIMO system greatly reduces the overhead and complexity associated with physical layer adaptation, where signal processing can play a role. There are many other discussions including signal detection \cite{otfs-signal} , modulator design \cite{otfs-design}, interference cancellation \cite{otfs-Interference}, and  performance analysis \cite{otfs-performance}.

In this part, we  focus on  the \emph{angle-delay-Doppler} 3D structured
sparsity of the OTFS channel in massive MIMO systems.
The downlink channel $\bm {\mathcal H}$ of the OTFS massive MIMO is a 3D tensor of dimension $N_e\times N_o\times M$, whose $(n_e, n_o,m)$th entry ($n_e \in \{1,2,\cdots,N_e\}$, $n_o \in\{1,2,\cdots,N_o\}$, and $m \in\{1,2,\cdots,M\}$) is given by \cite{otfs-3D-channel}
\begin{align}\label{OTFS1}
\bm {\mathcal H}_{n_e,n_o,m}\!=&\!\sum_{p=1}^{P}\beta_p \Psi(n_eT_s\!-\!\tau_p)\Upsilon_{\!N_o}(n_o\!\!-\!\nu_p N_o\triangle T\!-\!\!N_o/2\!-\!1\!)\notag\\
&\times \Upsilon_{M}(m-M(1+\sin\theta_p)/2\!-1\!),
\end{align}
where $T_s$ is the system sampling interval, $\Psi(\tau)$ is the band-limited pulse shaping filter (e.g., the Hanning window), $\Upsilon_{M}(x)\triangleq\sum_{m=1}^{M}e^{j2\pi\frac{m-1}{M}}$,  and $n_e, n_o,m$ corresponding  to the delay, Doppler, and angle index, respectively.
\begin{figure}[t]
  \centering
  \vspace{-0.5cm}
  \includegraphics[width=80mm]{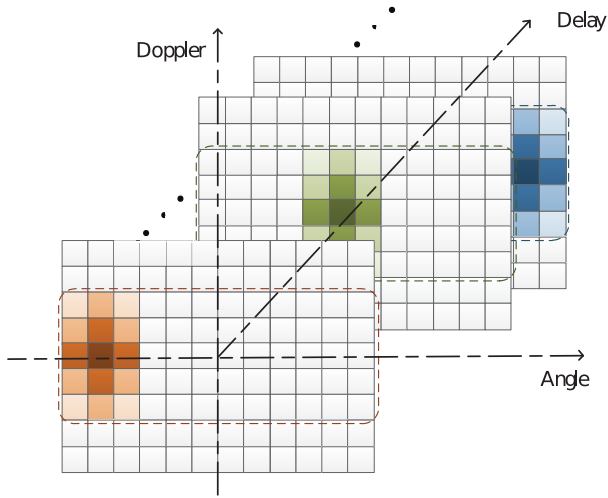}
  \caption{Angle-delay-Doppler 3D channel, which is burst-sparse along
the angle dimension, sparse along the delay dimension, and  block-sparse along the Doppler dimension \cite{otfs-3D-channel}.}
  \label{fig-otfs}
\end{figure}
The function $\Upsilon_{M}(x)$  has the property $| \Upsilon_{M}(x)| \approx 0$ when $|x|\gg 1$.
Then $H_{n_e,n_o,m}$  has the dominant elements  only if $n_e=\tau_p/T_s$,
$n_o=\nu_p N_o\triangle T +N_o/2+\!1$, and $m=M(1+\sin\theta_p)/2\!+1$, as shown in the Fig. \ref{fig-otfs} \cite{otfs-3D-channel}.
Since the number of significant  propagation paths is typically  limited, the 3D time-variant channel is sparse along the delay dimension.
Also,  the Doppler frequency of a path is usually
much smaller than the system bandwidth, the 3D channel
is block-sparse along the Doppler dimension,
Besides, with the assumption  of narrow AS, the 3D channel is burst-sparse along the angle
dimension \cite{otfs-burst-sparsity}. The lengths of non-zero bursts can be
regarded as constant, but the start position of each nonzero
burst is unknown, as shown in  Fig. \ref{fig-otfs}. This 3D sparse property indicates that the high dimension channel can be  efficiently estimated via limited  channel parameters with low overhead.
However, the existing OTFS works estimated the channel parameters via on-grid methods where  the power leakage effect still exists. Hence, future researches are still demanded for developing off-grid parametric algorithms.
\section{Multiple Access}
The key advantage of massive MIMO is to tremendously enhance the system capacity by simultaneously serving multiple users in the same frequency and time resource.  With the knowledge of CSI, the conventional 1G to 4G mobile communication networks generally assign orthogonal time, frequency, or spatial resource for different users. Nevertheless, these multiple access schemes can be re-visited under the  massive MIMO configuration with the aid of array signal processing.
\vspace{-2mm}
\subsection{Angle Division Multiple Access (ADMA)}
\begin{figure}[t]
  \centering
  \vspace{-0.5cm}
  \includegraphics[width=80mm]{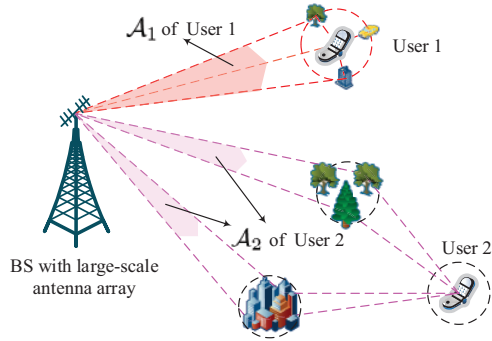}
  \caption{ADMA for users with non-overlapping angular spreads. }
  \label{ADMA}
\end{figure}
From \eqref{htf-narrow-vector}, it can be observed that the spatial domain channel and angle domain parameters  formulate a Fourier transform pair, especially when $M$ goes to infinity.  For conventional MIMO case, $M$ is too small to discriminate AoAs, such that one can only rely on the spatial channel $\bm h$ to design multiple access schemes. However, with massive array and sparse scenario, as explained previously, the AoAs can be well extracted and could be used for discriminating  users too. It has been implicitly shown in \cite{lowrank2,JSDM} that the orthogonal eigen-space based multiple access scheme is equivalent to using non-overlapping angular spread to distinguish different users. In \cite{hx-dft,linhan-dft}, non-overlapping angular spread were explicitly exploited for multiple users access  and formulated the so-called angle division multiple access (ADMA), as shown in Fig. \ref{ADMA}. It was actually proved in \cite{hx-dft,linhan-dft} that users with non-overlapping angular spread would result in orthogonal spatial channels  under massive MIMO configurations.

However, having orthogonal CSI from two users does not necessarily mean that the angular spreads of the users are non-overlapping, especially for conventional small MIMO. Actually, the conventional spatial domain multiple access (SDMA) is  universal for any number of antennas while ADMA can only be applied when AoAs of users are distinguishable, say in massive MIMO and sparse scenario. Nevertheless, the advantage of ADMA is drawn from the frequency insensitivity of the angular parameters, i.e.,  even if the frequency changes (uplink and downlink in FDD systems), the multi-users access designed from ADMA is still applicable.
Moreover, the ADMA would link the users' transmission with their individual  locations such that many geometric theories could be utilized to design the multi-user scheduling, even for multiple BS stations or the cell-free case \cite{hx-full}.
An example that adopts two BS and geometric theory to eliminate the blockage problem is shown in Fig. \ref{space}.
\begin{figure}[t]
  \centering
  \vspace{-0.5cm}
  \includegraphics[scale=0.5]{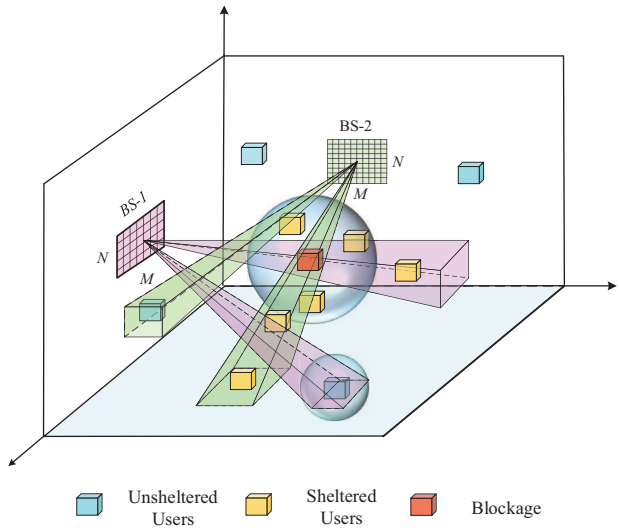}
  \caption{Illustration of the ADMA to eliminate the blockage problem in 3D
full space. }
  \label{space}
\end{figure}
\vspace{-3mm}
\subsection{Delay Division Multiple Access (DDMA)}
Similarly, the frequency domain channel and the delay  domain parameters  formulate another Fourier transform pair. For the conventional narrowband case, the bandwidth $W$ is too small to discriminate multi-path delays, such that one can only rely on the frequency domain orthogonality to design multiple access schemes. However, under the broadband transmission (mmWave) and sparse scenario,  the delay $\tau_p$ can be well extracted from array signal processing techniques. It can be easily shown that users with non-overlapping delays under broadband configuration will exhibit orthogonal frequency domain channels, which can  be used for distinguishing  users. The corresponding multiple access  scheme is named as delay division multiple access (DDMA). Interestingly, delay parameters can be artificially controlled as compared to AoA parameters. For example, even if two users have overlapping delay parameters, one user can purposely postpone its transmission such that the two users would have non-overlapping delays,  as shown in Fig. \ref{DDMA}.
However, the orthogonal frequency domain channels from two users do not necessarily mean that the delay of the users are non-overlapping, especially for narrowband systems. Actually, the advantage of the conventional frequency domain multiple access (FDMA) is its universal applicability for any bandwidth while DDMA can only be applied when the delays of different users are distinguishable.   Nevertheless, the advantage of DDMA is also drawn from the frequency insensitivity of the delay parameters, which says that even if the frequency changes (uplink and downlink in FDD systems),  the multi-users access designed from DDMA is still applicable.
\subsection{Doppler Division Multiple Access (DoDMA)}
Lastly, it can be observed that the time domain channel and the Doppler domain parameters formulate the third Fourier transform pair. Hence, we could theoretically  predict that it is possible to extract the Doppler parameters of different users and then discriminate users with non-overlapping Doppler  parameters. For example, for wideband communications in a  sparse environment, the symbol duration $T$ is very small such that a large amount of  symbols (corresponding to large array for AoAs or large bandwidth for delays) can be observed in a short time, during which period the Doppler parameters remain the same. In this case, the Doppler parameters $\nu_{p}$ can be well extracted and used for discriminating  users too. The resultant scheme is name as Doppler  division multiple access (DoDMA) or velocity division multiple access (VDMA). It can be easily showed that users with non-overlapping Doppler shifts will result into orthogonal time domain channel.
\begin{figure}[t]
  \centering
  \vspace{-0.4cm}
  \includegraphics[width=90mm]{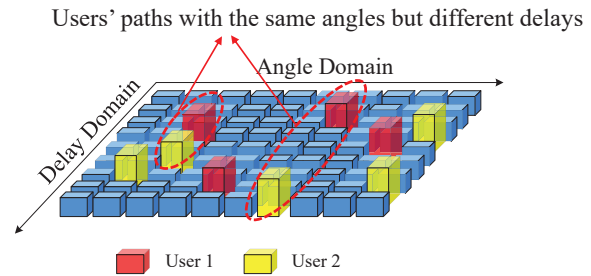}
  \caption{DDMA for users with  non-overlapping delays.}
  \label{DDMA}
\end{figure}
However, orthogonal time domain channels from two users do not necessarily mean that the Doppler shifts of the users are non-overlapping. Actually, the advantage of the conventional time domain multiple access (TDMA) is  universal for any bandwidth while DoDMA can only be applied when Doppler shifts of users are distinguishable.  Nevertheless, the advantage of DoDMA is also drawn from the frequency insensitivity (with a known scaling factor) of the Doppler parameters, which says that even if the frequency changes (uplink and downlink in FDD systems), the multi-users access  designed can be adapted to the new frequency band easily.  However, DoDMA may be difficult to implement in practice because once users are moving, their Doppler parameters would be  varying and may cause a frequent update of the multiple access strategy.
\begin{figure}[t]
  \centering
  \vspace{-0.5cm}
  \includegraphics[scale=0.5]{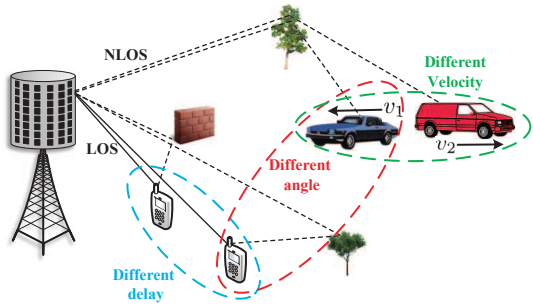}
  \caption{Massive MIMO with PDMA.}
  \label{PDMA}
\end{figure}
\vspace{-3mm}
\subsection{Path  Division Multiple Access (PDMA)}
The previously introduced ADMA, DDMA, and DoDMA are all based on the physical parameters of the multi-paths, which together can be named as path division multiple access (PDMA).\footnote{The term PDMA has been used in lens array \cite{PDMA-ZhangRui} but in fact only refers to angle domain.}
A diagram  to represent different multiple access schemes as well as their relationships is shown in Fig. \ref{PDMA}.

The logical relationship between the aforementioned six multiple access techniques is shown in Fig. \ref{6D}.
According to different situations, one can flexibly choose various combinations of the above X-DMA to maximize the utilization of the time/frequency/spatial resource. For example, with massive MIMO and narrowband transmission where AoAs can be discriminated but the delays cannot, one can choose  ADMA combined with FDMA for multi-user access.
Moreover, for multi-user OTFS massive MIMO systems, if angle/delay/ Doppler are utilized  simultaneously for multiple access, then it is  a kind of PDMA.

Nevertheless, it must be pointed out that introducing ADMA, DDMA, and DoDMA with the aid of array signal processing do not mean we can achieve three completely new degree of freedoms. For example, SDMA and ADMA are inherently from the same Fourier pair and hence it is not possible to apply SDMA and ADMA simultaneously.
\section{Artificial Intelligence Methods for Massive MIMO}
In recent years, artificial intelligence (AI), e.g., machine learning (ML) and deep learning (DL), have achieved great success in the fields of computer vision, natural language processing, speech recognition, etc. \cite {DL-survey1}.
Researchers in the field of wireless communications  are eager to apply AI technologies in all levels of the wireless systems that could  create \emph{intelligent communications} to meet various demands for 5G beyond \cite {DL-survey2}.

Inspired by the recent advances in ML/DL paradigm, ML/DL-based wireless communication techniques have aroused considerable interest among the academic and industrial communities \cite{DL-survey1,DL-survey2,DL-survey3}.
In this section, we investigate  ML/DL-based signal processing techniques for parameter-based  channels in massive MIMO systems, including channel estimation,  beamforming, etc.
\begin{figure}[t]
  \centering
  \hspace{-7mm}
  \vspace{-5mm}
  \includegraphics[width=90mm]{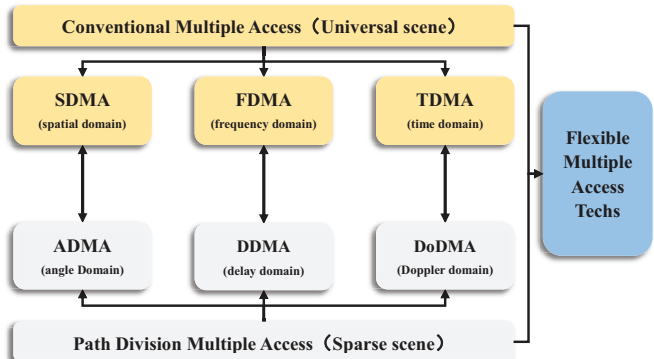}
  \caption{PDMA and traditional multiple access technologies to construct X-DMA.}
  \label{6D}
\end{figure}
\vspace{-3mm}
\subsection{Learning Based Channel Estimation}
In \cite{DL-Channel1}, the authors exploited a learned denoising-based approximate message passing
(LDAMP) network to solve the channel estimation problem, which was the first attempt to use DL technology for beamspace channel estimation.
The LDAMP  network regards the channel matrix as a 2D natural image and  incorporate the denoising convolutional neural network (DnCNN) into the iterative signal recovery algorithm for channel estimation.
With large number of channel matrices being training data, the LDAMP can be applied to a variety of selection networks.
In order to estimate AoA with high resolution, Huang {\emph{et al.} \cite{DL-Channel2} utilized a deep neural network (DNN) that integrated the massive MIMO into DL.
The DNN was employed to conduct off-line learning and online learning procedures, which learned the statistics of the wireless channel and the spatial structures in the angle domain.
In addition, an online DNN based channel estimation algorithm for doubly selective fading channels was proposed in \cite{DL-Channel3}.  With properly selected inputs, the DNN can not only exploit the features of channel variation from previous channel estimations but also extract additional features from pilots and received signals \cite{DL-Channel3}.
\vspace{-3mm}
\subsection{Learning Based Beamforming}
To solve the analog beam selection problem for hybrid beamforming in mmWave massive MIMO systems,  Long {\emph{et al.} \cite{DL-BF1} proposed a data-driven ML solution by resorting to support vector machine (SVM).
In order to reduce the complexity of classification, the AoAs and AoDs of  multi-paths  were used as entries of feature vectors, and a large number of samples from the mmWave channels were collected as the training data.
Then,  Ant$\acute{\text{o}}$n-Haro and  Mestre  \cite{DL-BF2} exploited  AoA information and learning  approaches to perform beam selection in  hybrid mmWave communication systems.
To compare with  different degrees of complexity/sophistication, \cite{DL-BF2} adopted three leaning methods including two ML approaches, SVM,  $k$-nearest neighbors, and one DL approach, the multi-layer perceptron.
The beam selection task was finally changed into a multi-class classification problem and was solved via those three schemes.
Meanwhile, Klautau {\emph{et al.}  \cite{DL-BF3} investigated the performance of several ML schemes (SVM, Adaboost, Decision trees, Random forests) but also DNN and reinforcement learning for beam  selection techniques on vehicle-to-infrastructure using mmWave.
\vspace{-2mm}
\subsection{Prospects of Learning Based Methods}
From the brief introduction, it can be seen that the trial of learning networks utilized in parametric massive MIMO systems  can achieve appealing performance in channel estimation,  beamforming, etc.
Many other extensions are also attracting, like CSI feedback/reconstruction \cite{DL-CSI2}, signal detection, channel coding/decoding, end-to-end communications systems (see \cite{DL-survey1,DL-survey2,DL-survey3} for more suggested future works).
Based on the previous sections, we firmly believe that
these AI methods could be applied from array signal processing viewpoints and would achieve all the advantages discussed preciously.
\section{Conclusions and Future Works}
\begin{figure}[t]
  \centering
  \vspace{-0.5cm}
  \includegraphics[width=90mm]{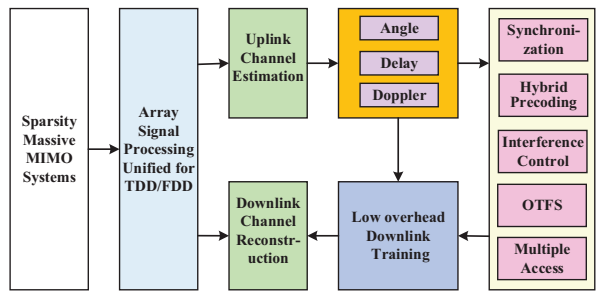}
  \caption{The structure of array signal processing techniques for  massive MIMO sparsity channels.}
  \label{system}
\end{figure}
The purpose of this paper is to provide an overview on  array signal processing techniques that have been successfully applied in the context of massive MIMO systems. These methods rely on sparse and low-rank signal models, for which the physical parameters of channels that characterize the channels responses, i.e., angle, multi-path delay, and Doppler can be well estimated with large array and broad bandwidth.  Moreover, we especially point out three new multiple access schemes as well as their relationships to the conventional TDMA FDMA, and SDMA.

The whole principle of applying array signal processing for massive MIMO system is summarized in Fig. \ref{system}, and the key advantages of applying array signal processing are summarized here:
\begin{itemize}
\item The number of the unknown parameters can be reduced such that the training overhead can be greatly shortened, and the estimation accuracy is
higher than the conventional approaches that simply assume channel sparsity;
\item Angle, delay, and Doppler parameters are widely frequency insensitive and the so designed transceiver techniques  like channel estimation, synchronization, precoding, and user scheduling, etc., can be well adapted from one frequency band to another.
\end{itemize}

Besides, with array signal processing, one can deeply understand the RF communications rationale and explain the effect of spatial- and frequency-wideband effect that was previously ignored from communications viewpoint. It is then not difficult to realize that most existing algorithms were designed only with frequency- and spatial- narrowband channel models.
Extensions to wideband channels in many cases are important and necessary, especially for mmWave, Terahertz massive MIMO systems and  extremely large MIMO systems.

Moreover, to fully utilize the array signal processing techniques in wireless communications, the problems faced by the traditional array signal processing should  also be studied. For example, the parameter-based channel models discussed in previous sections are  sensitive to array response errors and how to calibrate large scale array in FDD systems is still an open problem \cite{caribation}.
Further works are needed to investigate the areas of array signal processing like target tracking, array positioning,  array shape design, array calibration, etc., that can all be re-shaped into wireless communications. Though we cannot spend more space to illustrate all these works, we hope this short overview paper can give a clear picture of why and how the array signal processing can be merged into wireless communications. We would also like to encourage researchers from both areas to devote more efforts into this promising direction.

\end{document}